\documentclass[preprint, english]{revtex4-1}
\usepackage[T1]{fontenc}
\usepackage[utf8]{luainputenc}
\usepackage{amsmath,amsfonts,amsthm,bm}
\usepackage{braket}
\usepackage{amssymb}
\usepackage{graphicx}
\usepackage{esint}
\usepackage{babel}

\begin{document}
	
	\title{A Modular Operator Approach to Entanglement of Causally Closed Regions}
	
	\author{Cosmo Gallaro}
	\author{Rupak Chatterjee}
	\email{corresponding author: Rupak.Chatterjee@Stevens.edu}
	
	\affiliation{Center for Quantum Science and Engineering\\
		Department of Physics, Stevens Institute of Technology, Castle Point on the Hudson, Hoboken, NJ 07030}
	
	\begin{abstract}
	Quantum entanglement is shown for causally separated regions along the radial direction by using a conformal quantum mechanical correspondence with conformal radial Killing fields of causal diamonds in Minkowski space. In particular, the theory of local von Neumann algebras and Tomita Takesaki modular operators is applied in the entanglement structure of causal diamonds in conformal quantum mechanics. The entanglement of local states in their respective causal regions is shown through the measures of concurrence and entanglement entropy using the Tomita Takesaki modular conjugation operator. A holographic entropy formula is derived for the conformal quantum mechanics causal diamond correspondence. A new connection is made between the thermal time flow defined by the modular group of automorphisms to the physical time flow in a causal diamond via the aforementioned correspondence. The thermal interpretation of these results via two-point thermal Green's functions and modular group flow supports the idea of a possible emergent theory of spacetime. 
	\end{abstract}
	
	\maketitle
	
	\section{Introduction}
	Conformal symmetry is a local scale symmetry that keeps the metric tensor of spacetime invariant up to a spacetime dependent scale, $ds^2 = \Omega(x) g_{\mu \nu}dx^{\mu} dx^{\nu}$. Physically, this means that a system is $locally$ invariant under the change of observation scale while leaving spacetime angles invariant. Indeed it is an extension of the Poincare group possessing 15 degrees of freedom, the extra five being four special conformal transformations and one dilation. This class of symmetry has been previously used to study critical phenomena such as phase transitions where the diverging correlation lengths indicate scale invariance. However, the symmetry need not only be applied to systems of infinite degrees of freedom. Conformal invariance in a quantum mechanical system was first investigated in \cite{Alfaro1976} and subsequently, with the interest in conformal field theories, seen as formally equivalent to a conformal quantum field theory in $(0+1)$ dimensions \cite{Boozer2007}. The conformal invariance analogy in a quantum mechanical system is made with respect to a conformal algebra rather than a metric tensor as in classical relativity.
	
	A $causal$ $diamond$ is some finite region of spacetime that encompasses the past, present and future of a particles spacetime world-line, whose causal geometric structure emerges from the collective emanation of light from the said region. Such a region, and the energy within it, could be interpreted as an observer with a finite lifetime \cite{Unruh1989}. In  \cite{Arzano2020}, Arzano made a connection between the conformal radial symmetries of causal diamonds in Minkowski space-time and time evolution in conformal quantum mechanics. Specifically, the two point function of conformal quantum mechanics was related to a two point function for an observer in Minkowski space, as well as the size of a causal diamond to a temperature parameter. A deep relationship between the generators of the radial conformal Killing vector field in Minkowski space, which describes the conformal symmetry of the radial dimension of space, and the generators of the conformal symmetry in quantum mechanics was found.
	
	Here, we investigate the entanglement properties within this correspondence by considering causal diamonds which are separated by a space-like distance in Minkowski space. In particular, we construct a Tomita-Takesaki modular operator formalism within a von Neumann algebra structure for two causal diamonds. This allows us to discuss the relation between conformal radial Killing fields of two causal diamonds in Minkowski space and their entanglement. Specifically, we use the modular conjugation operator to express Wooter's concurrence for arbitrary states in two causally separated regions. The construction of concurrence in terms of the modular conjugation operator was first discussed using a super-symmetric quantum mechanical system in \cite{Chatterjee2021}. It is known in holography theory that bulk causality can emerge from a boundary conformal field theory via a definition of modular flow and its entanglement between distant regions \cite{Maldacena1999, Maldacena2003}. Conformal quantum mechanics is an excellent model to better understand the emergence of bulk causality from a (0+1) conformal field theory and arising entanglement due to separation of causal regions through use of von Neumann algebras and Tomita-Takesaki modular theory. An entropy formula is derived which is found to be proportional to the length of the causal diamond with a corrective logarithmic term, similar to that found in \cite{Carlip2000}. The $(0+1)$ conformal field theory is mapped to a $(1+1)$ causal physical system, leading to the concept of a "line density" of entanglement entropy, similar to the holographic area density of Bekenstein and Hawking. The framework developed in this paper may further elucidate the holographic emergence of bulk causality. The problem that we solve is constructing local von Neumann algebras for causal diamonds using the correspondence with conformal quantum mechanics. Furthermore, this allows us to define Tomita-Takesaki modular operators and provide holographic entanglement entropy for a cyclic vector state of the von Neumann algebra representing the radial dimension of a causal diamond. Finally, we then comment on the interpretation of the model and its possible origination in thermality.
	
	\section{The Correspondence between Classical and Quantum Symmetry}

	de Alfaro et al. \cite{Alfaro1976} were the first to deduce from group theoretic principles the algebra of quantum mechanical operators that correspond to the algebra of infinitesimal generators of conformal transformations. They do this by examining projective M\"{o}bius transformations for a time variable $t$ in a $(0+1)$ conformal field theory. Using Heisenberg's equations of motion, they are able to match the conformal generators to their unitary time evolution operators for each transformation generator. It it thus interpreted that the theory is performing projective conformal transformations over time. The conformal algebra is infinite in two dimensions making any problem in 2D exactly solvable. Though we have $(0+1)$ for this model, the theory is still exactly solvable since the generators of conformal transformations are constants of motion. The Lagrangian for a quantum mechanical theory can be written as 
	\begin{equation}
	    L = \dfrac{1}{2}\left(\partial_t Q - \dfrac{g}{Q^2}\right)
	\end{equation}
	for some Hermitian quantum field $Q(t)$ that only depends on time, hence a (0+1) field theory. It is the action, $A = \int L \; dt$, that is invariant under conformal transformations. The transformations of the conformal group can be found via an infinitesimal analysis \cite{Alfaro1976}. These transformations include translations generated by $H = \dfrac{d}{dt}$, dilations generated by $D = - t \dfrac{d}{dt}$, and special conformal transformations generated by $K = - t^2 \dfrac{d}{dt}$. The generators of time evolution in conformal quantum mechanics thus form the $\mathfrak{s}l(2,\mathbb{R})$ algebra \cite{Alfaro1976},
	\begin{equation}
		[H,P] = iH ,\; [K,D] = -iK , \; [H,K] = 2iD
	\end{equation}
	Furthermore, a combination of these generators yields the observable quantity, 
	
	\begin{equation}
	    	G = uH + vD + wK.
    \label{G}\end{equation}  
    This generator leaves the action of the quantum field invariant and is a constant of motion. This means that one could choose any one of these conformal generators to study time evolution, though they have different properties.  
    
    Consider now (3+1) dimensional Minkowski space associated with some symmetries. A conformal radial Killing field is a radial vector field such that the Lie derivative acting on the metric is proportional to the metric itself, $\mathcal{L}_\chi \eta_{\mu \nu}$  $\alpha$  $\eta_{\mu \nu}$, demonstrating conformal invariance of the spacetime. Such a Killing field, $\chi$, can be written as a combination of three conformal generators:
	
	\begin{equation}
	    \begin{array}{c}
		    \chi = \left\{a(t^2 + r^2) + bt + c\right\}\partial_t + r(2at +b)\partial_r \\
		    = aP_0 + bD_0 +cK_0
		\end{array}
	\end{equation}
    Arzano \cite{Arzano2020} obtains the relationship between generators of conformal quantum mechanics and the conformal radial killing field as: 
    \begin{equation}
        \begin{array}{c}
		    G = i\chi \\\\
		    H=iP_0, \;D=iD_0, \; K=iK_0 \\\\
		    a = u, \; b =v, \; c = w
		\end{array}
	\end{equation}
	That is, the generators of conformal symmetries in quantum mechanics, $H$, $D$, and $K$ are directly related to the conformal radial Killing vector field that describes the symmetry of the radial dimension of a causal diamond in Minkowski space. The generators of conformal radial symmetries of a causal diamond in Minkowski space are analytically continued into the complex plane and are expressed with only quantum mechanical time evolution. Therefore, the classical conformal symmetry of the radial and temporal coordinate in Minkowski space, $(r,t)$, appear to emerge from generators of  conformal symmetry in quantum mechanics \cite{Chamon2011}. In other words, a (1+1) conformal spacetime can be described with a (0+1) conformal quantum theory. The first theory of this kind was the AdS/CFT correspondence of Maldacena \cite{Maldacena1999}.

	There are classes of Hamiltonians that are characterized the determinant of the generator $G$, as it is invariant under general transformations. The Hamiltonian corresponding to a negative determinant of $G$ is a \textit{rotational} Hamiltonian that generates elliptical compact rotations. For now, we recognize that this operator is convenient to use for calculations as its basis is discrete and complete, and we thus proceed with the model in this rotational basis. The operator $R$ is given as, 
	
	\begin{equation}
		\begin{array}{cc}
			R = \dfrac{1}{2}\left(\alpha H +\dfrac{K}{\alpha}\right)
		\end{array}
	\end{equation}
    and generates elliptical rotations. The other classes of Hamiltonians are $H$ and $K$ generating parabolic transformations around the light-like axis when the determinant is zero, while $D$ and $S=\dfrac{1}{2}\left(\alpha H -\dfrac{K}{\alpha}\right)$  generate hyperbolic transformations (Lorentz boosts) when the determinant is positive. Note that the number $\alpha$ has units of time  \cite{Alfaro1976,Arzano2020}.
    
	The connection between the quantum conformal generators and the Killing field is illustrated through the fact that their classification depends upon the determinant of the Casimir operator of the conformal group as explained in \cite{Arzano2020}. This connection highlights how the three classes of 'Hamiltonians' generate different types of time evolution. Indeed, one can express the three 'Hamiltonians' $H$, $R$, and $S$ in terms of three different time coordinates,
	
	\begin{equation}
			H = i\partial_t, \;
			R =i\partial_T, \;
			S=i\partial_{\tau},
	\end{equation}
where the times are related to each other through
	
	\begin{equation} 
		t = \alpha tan(T/2) = \alpha tanh(\tau /2)
	\end{equation}
	
	Upon quantization, the rotation $operator$ $R$ has eigenstates and a corresponding rotational vacuum $\ket{n=0} $ such that \cite{Alfaro1976}
	
	\begin{equation}
		\begin{array}{c}
			R\ket{n} = r_n \ket{n} \\\\
			L_{\pm}\ket{n} = \sqrt{r_n(r_N\pm1)-r_0(r_0-1)}\ket{n\pm1}  \\\\
			R\ket{n=0} = r_0\ket{n=0} \\\\
			r_n  = r_0 + n,\; n=0,1,2,...
		\end{array}
	\end{equation}	
	with orthonormal eigenstates. One can introduce creation and annihilation operators for the rotational eigenstates leading to 
	\begin{equation}
	\begin{array}{c}
    R = \frac{1}{2}(a^{\dagger}a + \frac{1}{2})  \rightarrow
    r_n = \frac{1}{2}(n+\frac{1}{2}) \\ \\
    L_+ = \frac{1}{2} a^{\dagger2}\\ \\
    L_- = \frac{1}{2}a^2
	\end{array}
	\end{equation}
    where we identify $L_+ $ and  $L_-$ as raising and lowering operators. 
	
	Furthermore, one can write an expression for raising and lowering operators of rotations,
	
	\begin{equation}
		L_{\pm} = \frac{1}{2}\left(\alpha H -\frac{K}{\alpha}\right) \pm iD
	\end{equation}
	to realize the algebra in the Cartan sub-basis,
	
	\begin{equation}
		[R,L_{\pm}] = \pm L_{\pm}, \; [L_-,L_+] = 2R
	\end{equation}
thus allowing us to define a scaling dimension and spin for the generators of dilations and rotations in conformal quantum mechanics by observing the conformal weights of the two point functions \cite{Henkel1999}. We can identify a relationship between the conformal radial Killing field in Minkowski space and the quantum rotational generator $R$ by noticing that the generator $R = i\partial_T$ can also be written as 
	
	\begin{equation} 
		R = \dfrac{1}{2\alpha} (\alpha^2 H + K) = \dfrac{i}{2\alpha} \left[ (t^2 + r^2 + \alpha^2) \partial_t + +2\;t\;r\partial_r \right] = i\chi 
	\end{equation}
 with $a=\dfrac{1}{2\alpha}$, $b=0$ and $c=\dfrac{\alpha}{2}$ using the relations $H=iP_0$ and $K=iK_0$. The parameter $\Delta = b^2 - 4ac$ determines the causal character of the radial conformal Killing field. For our case, $\Delta = -1$ and thus the causal character is time-like everywhere. It is also interesting to view the Killing field in terms of the $(0+1)$ quantum field $Q(t)$. Using the expressions for the generators in terms of the field $Q(t)$ found in \cite{Alfaro1976}, we obtain the relationship between $R$ and $Q(t)$, 
	
	\begin{equation}
			R = \frac{1}{2}\left(\alpha (\partial_t Q^2 + g/Q^2) + \frac{1}{\alpha} \left[t^2 / 2  (\partial_t Q^2 +g/Q^2) 
			- \frac{1}{2}t(Q \partial_t Q + \partial_t Q Q) + \frac{1}{2}Q^2\right]\right)
	\end{equation}
and we can express the conformal radial Killing field for a causal diamond in Minkowski space in terms of the quantum field $Q(t)$,
	
	\begin{equation} 
	    \begin{array}{c}
    		\chi  =  \dfrac{-i}{2}\left(\alpha (\partial_t Q^2 + \dfrac{g}{Q^2}) + \dfrac{1}{\alpha} \left[t^2 / 2  (\partial_t Q^2 +\dfrac{g}{Q^2}) 
			- \dfrac{1}{2}t(Q \; \partial_t Q + \partial_t Q \;  Q) + \dfrac{1}{2}Q^2\right]\right)
    	\end{array}
	\label{killingfieldquantumfield} \end{equation}
 providing a relationship between the classical radial conformal Killing field of a causal diamond and the $(0+1)$ dimensional quantum field. From the relation $-iR = \chi = \partial_{T}$ one can adopt diamond coordinates as
	\begin{equation} 
		t =\alpha \frac{sin(T)}{cos(x) + cos(T)}  , \;\; r = \alpha \frac{sin(x)}{cos(x) + cos(T)} 
	\end{equation}
and transform the typical Minkowski line element $ds^2=-dt^2+dr^2+r^2 d \Omega^2 = -dt^2+dr^2+r^2 (d\theta^2+\sin^2\theta d\phi^2)$ into
	\begin{equation} 
		ds^2 = \frac{\alpha^2}{(cos(T) + cos(x))^2} (-dT^2 + dx^2 + sin^2xd\Omega^2) 
	\label{ds}\end{equation}
sufficing as a $conformal$ transformation in the radial and time dimensions only. That invites one to think of this as a holographic theory from which $(1+1)$ spacetime can be described as a $(0+1)$ conformal field theory.

Using the integral curves of conformal radial Killing fields found in \cite{Herrero1999}, we set $a=\dfrac{1}{2\alpha}$, $b=0$, and $c = \dfrac{\alpha}{2}$ for the case of $-iR = \chi$ to obtain the one-parameter family of world-lines,
\begin{equation}
    \begin{array}{c}
        \dfrac{1}{2\alpha}(t^2 - r^2) -\omega r + \dfrac{\alpha}{2} =0 \\\\
        t^2 - (r-\alpha\omega)^2 = -\alpha^2(1+\omega^2)
    \end{array}
\end{equation}
for some parameter $\omega$ giving a family of hyperbola in the $(t,r)$ Minkowski space. From these curves, we note the corresponding acceleration of an observer in these coordinates as \cite{Herrero1999},
\begin{equation}
    a =\dfrac{1}{\alpha \sqrt{\omega^2 +1}} = \dfrac{sin(x)}{\alpha}
\end{equation}
for the parameter $\omega = \dfrac{1}{tan(x)}$. This corresponds to hyperbolic relativistic motion. The acceleration is zero when passing through $x=0$ where we can also notice that 
\begin{equation}
    t = \alpha \dfrac{sin(T)}{1+cos(T)} = \alpha tan(T/2)
\end{equation}
relating the diamond time $t$ to the time variable of the conformal quantum rotation operator $T$. 
As is well known, there is a deep connection between temperature and acceleration, as illustrated through the Unruh/Hawking effect \cite{Unruh1976, Hawking1975}. There is also a direct relationship between the acceleration of a massless scalar field along the world-line of a static observer in a causal diamond and the temperature of the primary conformal two point function of conformal quantum mechanics \cite{Arzano2020, Anninos2012, Nakayama2012} with the appropriate conformal weight. The length of this causal diamond is $2\alpha$ \cite{Unruh1989} as seen from the line element (\ref{ds}), and is inversely proportional to the temperature found in the thermal Greens function as $T=\dfrac{1}{\pi \alpha}$. It was in fact mentioned in \cite{Unruh1989, Arzano2020, Anninos2012} that $2\alpha$ is the lifetime of a static observer in Minkowski space. The construction in terms of the operator $R$ was built in order to highlight the connections between two separate causal diamonds via the correspondence reviewed in this section. Indeed, it is the operator $R$ whose eigenbasis we use to define the cyclic vector for Tomita-Takesaki modular operators in the next section.

\section{Causality in Conformal Quantum Mechanics: Modular Operators}

    We begin by imagining that there are two causal diamonds in Minkowski space that are separated by some spacelike distance such that there is no causal relationship between observations in each region.  The system is illustrated in Figure 1. Each diamond has associated with their conformal radial symmetries a set of conformal quantum mechanical operators through the correspondence reviewed in the previous section. In this section, we show that a holomorphic representation of these quantum mechanical operators may be used to form von Neumann algebras of observables such that the second diamond's algebra of observables is shown to be the commutant of the first diamonds algebra. The conformal symmetries of each diamond is interpreted to emerge from the quantum generators $R$ of the kets $\ket{n}$ as reviewed in the previous section. 
	
	The relationship between these two systems of algebras, shown below, may be generated by Tomita-Takesaki modular operators from a standard form of von Neumann algebras (see the appendix and  \cite{Haag1996}). Proceeding with the GNS construction (see appendix), let $\mathcal{A}$ be a representation of a von Neumann algebra of bounded linear operators  on a Hilbert space $\mathcal{H}$, $\mathcal{A} \subset \mathcal{B(H)}$.  $\mathcal{A}$ is a $C^*$-algebra with a commutant $\mathcal{A}'$ (the set of elements in $\mathcal{B(H)}$ commuting with $\mathcal{A}$) such that $\mathcal{A}''=\mathcal{A}$ (the double commutant returns one to $\mathcal{A}$ as defined in the appendix). Let $|\Omega \rangle  \in \mathcal{H} $ be a separating and cyclic vector for the $C^*$-algebra $\mathcal{A}$ (the elements in  $\mathcal{A}$ can create a space dense in $\mathcal{H}$ by acting on $|\Omega \rangle  \in \mathcal{H} $). The Reeh-Schlieder theorem states that mathematically, the vacuum state of some quantum field theory is cylic $\mathcal{A}(\mathcal{O})\ket{\Omega}$ is dense in the whole Hilbert space $\mathcal{H}$ for any open region $\mathcal{O}$ in Minkowski space. Further, $\mathcal{A}(\mathcal{O}^*)\ket{\Omega}$ is also dense in $\mathcal{H}$, where $\mathcal{O}^*$ is the causal compliment to $\mathcal{O}$. Physically, this suggests that energy can be created or annihilated anywhere in the universe from anywhere in the universe, though  there are of course caveats to this statement. See Witten's review \cite{Witten2018} for an excellent discussion and proof. In the current theory, however, the vacuum is not cyclic as there is not an invariant vacuum state for $\mathfrak{s}l(2,\mathbb{R})$ \cite{Chamon2011} and so the Reeh-Schlieder theorem does not immediately follow. It is of great interest to us, however, to understand how the Reeh-Schlieder theorem can be analogously defined in a conformal quantum mechanics model.
	
	Tomita-Takesaki theory states that there exists an anti-linear map $ S: \mathcal{H} \rightarrow \mathcal{H},$ such that $ S A |\Omega \rangle  = A^* |\Omega \rangle, \forall A \in \mathcal{A} $. $S$ has a polar decomposition given by $S = J \Delta^{1/2} =  \Delta^{-1/2} J, \,\, \Delta = S^{*} S$ where the modular conjugation operator $J$ has the following properties (see appendix),
\begin{equation}
J \Delta ^{\frac{1}{2}} J = \Delta ^{-\frac{1}{2}}, \;\;\;
J^2 =I, \,\,\, J^{*} = J, \;\;\;
J |\Omega \rangle  = |\Omega \rangle,  \;\;\;
J \mathcal{A} J = \mathcal{A}' .
\end{equation}
One of the key feature for our purposes is that of the modular conjugation operator $J$ that takes an algebra $\mathcal{A}$ into its commutant $\mathcal{A}'$. 

Two causal diamonds separated by some spacelike distance will have local algebras of bounded operators acting on their respective local Hilbert spaces as define through the conformal quantum mechanical correspondence. To define some notion of causality, we demonstrate that the Hermitian operators corresponding to observables commute with the operators in the other causal diamond using Tomita-Takesaki modular theory. This defines the commutant of the algebra, and allows us to define local von Neumann Algebras in the two distant causal diamonds \cite{Haag1996}. While this is known in local algebraic quantum field theory, it is through the conformal quantum mechanical correspondence that we construct the commutant structure of local von Neumann algebras for the radial direction of causal diamonds. 

In the first causal diamond, we have local rotational quanta $\ket{n}$ described by a set of operators $a$ and $a^{\dagger}$. In the second diamond, we have local rotational quanta described by a different set of operators $b$ and $b^{\dagger}$ that will be shown to be related via involution. The algebraic structure of the double diamond system is given by,   
	
	\begin{equation} 
		\begin{array}{cc}
			\mathcal{H} = \mathcal{H}_a \otimes \mathcal{H}_b \\\\
			\mathcal{A}_a = \set{\mathcal{O}_a} \subset \mathcal{B} (\mathcal{H}_a) \\\\
			\mathcal{A}_b = \set{\mathcal{O}_b} \subset \mathcal{B} (\mathcal{H}_b)\\\\
			\mathcal{A} = \mathcal{A}_a \otimes \mathcal{A}_b
		\end{array}
	\end{equation}

\begin{figure}
\includegraphics[scale=0.45]{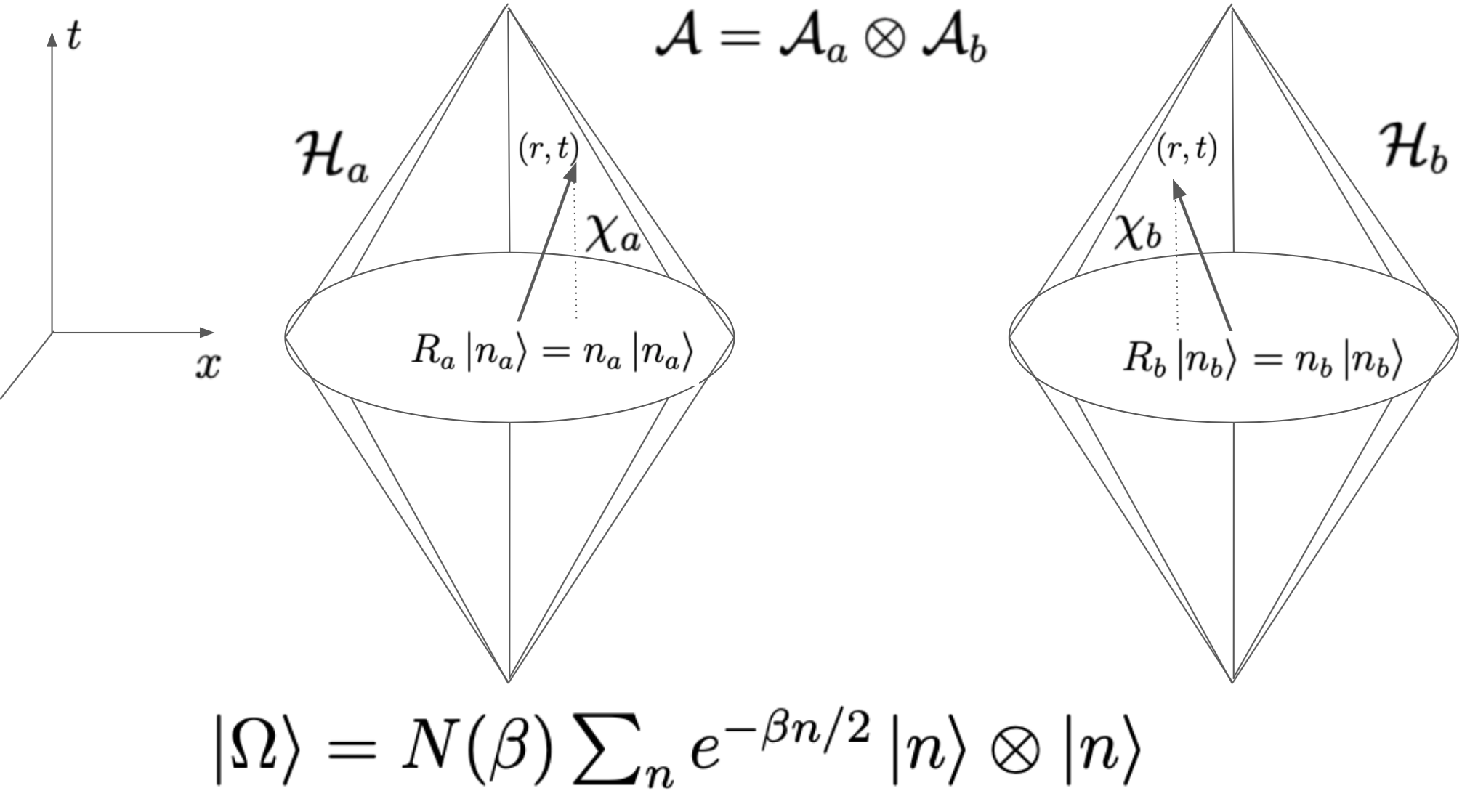}
\caption{Double Causal Diamond Structure}
\end{figure}
 
We place the operators $a$ and $a^{\dagger}$ in a holomorphic representation as in \cite{Chatterjee2021},
  \begin{equation}
      a = \dfrac{1}{\sqrt{2}}\left(z + \partial_{\bar{z}}\right) ,\; a^{\dagger} = \dfrac{1}{\sqrt{2}}\left(\bar{z} - \partial_{z}\right)
  \end{equation}
  such that $[a,a^{\dagger}] = 1$. We define the involution of the structure as transforming $a$ to $b$ such that 
	\begin{equation}
	    \begin{array}{c}
    	    a^* = \dfrac{1}{\sqrt{2}}\left(\bar{z} + \partial_{z}\right)  = b \\\\
    	    a^{\dagger*}  = \dfrac{1}{\sqrt{2}}\left(z - \partial_{\bar{z}}\right) = b^{\dagger}
    	\end{array}
	\end{equation}
	thus $[b,b^{\dagger}] = 1$ and therefore
	\begin{equation}
	    [a,b] = [a^{\dagger}, b^{\dagger}] = 0
	\end{equation}
    Then, $R_b = \dfrac{1}{2}\left(b^{\dagger}b+\dfrac{1}{2}\right)$. The modular conjugation operator $J$ will transform states from one diamond to another.
    
    The Tomita-Takesaki modular conjugation operator for this theory fulfills the identity for an arbitrary state, 
	
	\begin{equation}
	    \begin{array}{c}
	    J(a\otimes I)J \ket{n} \otimes \ket{m} =
	    J(a\otimes I)\ket{m} \otimes \ket{n} \\ \\
	    =J \sqrt{m}\ket{m-1 } \otimes \ket{n}
	    = \sqrt{m}\ket{n} \otimes \ket{m-1} \\ \\
	    =(I\otimes b)\ket{n} \otimes \ket{m}
	    \end{array}
	\end{equation}
	for the modular operator
	\begin{equation}
	    \Delta = e^{-\beta(N_a \otimes I  - I \otimes N_b)}
	\end{equation}
	and the physical state
	\begin{equation} 
	    \begin{array}{cc}
    		\ket{\Omega} = N(\beta) \sum_n e^{-\beta n/2}\ket{n} \otimes \ket{n} \\\\ 
    		= N(\beta) \sum_n \dfrac{e^{-\beta n/2}}{n!} (a^{\dagger})^n \ket{0} \otimes (b^{\dagger})^n \ket{0}
		\label{cyclic}\end{array}
	\end{equation}
	where $N(\beta) = (1-e^{-\beta})^{1/2}$, which is indeed cyclic and separating as there are no vanishing Schmidt coefficients \cite{Lashkari2019}. Further relationships are fulfilled,
	\begin{equation}
	   \begin{array}{c}
	    J\Delta^{1/2}(a \otimes I) \ket{\Omega}  =
	    Je^{-\frac{\beta}{2}(N_a \otimes I - I \otimes N_b)}(a \otimes I) \ket{\Omega} \\ \\
	   =Je^{-\frac{\beta}{2}(N_a \otimes I - I \otimes N_b)}(a \otimes I)N(\beta) \sum_n e^{-\beta n/2}\ket{n} \otimes \ket{n} \\ \\
	    =Je^{-\frac{\beta}{2}(N_a \otimes I - I \otimes N_b)} N(\beta) \sum_n e^{-\beta n/2}\sqrt{n} \ket{n-1} \otimes \ket{n} \\ \\
	    =J N(\beta) \sum_n e^{-\beta n/2}e^{-\frac{\beta}{2}(2n-1-2n)}\sqrt{n}\ket{n-1} \otimes \ket{n}  \\ \\
	    = N(\beta) \sum_n e^{-\beta n/2}e^{\frac{\beta}{2}} \sqrt{n} \ket{n} \otimes \ket{n-1} \\ \\
	    =N(\beta) \sum_n e^{-\beta n/2}e^{\frac{\beta}{2}} (I \otimes b) \ket{n} \otimes \ket{n}
	    = a^* \ket{\Omega}
	   \end{array}
	\end{equation}
	where the vector has been renormalized. The commutant structure ensures Einstein causality as the commutators vanish for operators in the two algebras and therefore generate the conformal radial symmetry of causally disconnected causal diamonds. Using these relations we can identify the commutant,
	
	\begin{equation} 
		\begin{array}{c}
			J R_a J = R_b 
		\end{array}
	\end{equation}
	The physical interpretation is made a little more clear by recalling the representation of $R$ in terms of some time coordinate $T$,
	
	\begin{equation} 
		R_a = i\partial_T
	\end{equation}
	such that
	
	\begin{equation}
		R_b = -i\partial_T
	\end{equation}
	which may hint at opposite flows of complex time evolution for space-like separated causal diamonds. Furthermore, since we also know that $R_a = i\chi_a$, we have $R_b = i\chi_b = -i\chi_a$. Clearly,
	
	\begin{equation} 
		[R_a, R_b] = [i\chi_a, i\chi_b ] = 0 
	\end{equation}
	 since two space-like separated casual diamonds will not have any causal influence on each others measurements or observations, making their respective symmetries commutative. Though there is non local interaction between these two diamonds, as we will see through the non-zero entanglement of the two diamonds which will be expressed below through the conformal quantum mechanical correspondence.

	 Within one causal diamond, there exists a bipartite structure as discussed in \cite{Arzano2021} of which this formalism may be applied. However, it is illuminating to consider one diamond and its causal compliment, of which a second diamond is a subset, as it is a common relationship to study in algebraic quantum field theory. Indeed, the identified commutant algebra is physically a subset of the causal compliment of the first algebra representing a causal diamond itself. In particular, the operators composing the commutant of conformal quantum mechanics also can be mapped to the radial conformal Killing field of another diamond. 

 The exponential argument $ N_a \otimes I - I \otimes N_b$ of the modular operator is serving as the modular Hamiltonian in this model generating rotations, as the quanta generated by the operator $R$ are indeed some type of abstract angular momentum, analogous to our familiar notion of quantum "spin" as being some angular momentum. The modular operator allows us to define a one parameter group of automorphisms on the algebra, $\mathcal{A}$, that generates the local time flow for the theory (see appendix),
	
	\begin{equation}
	    \alpha_t A = \Delta^{it}A\Delta^{-it} = e^{it\beta(N_a \otimes I - I \otimes N_b)}Ae^{-it\beta(N_a \otimes I - I\otimes  N_b)}
	\end{equation}
	experienced by observables on the algebras such that they will remain in such algebras. 
    Furthermore, due to the fact that our system contains finite degrees of freedom, we can define a suitable density matrix, $\rho  = Z^{-1}e^{-\beta(N_a \otimes I - I\otimes  N_b)}$ where $Z = Tr(e^{\beta(N_a \otimes I - I\otimes  N_b)})$ is the partition function. An algebraic state can thus be written for some operator $A \in \mathcal{A}$,
	
	\begin{equation}
	    \omega (A) = \mathrm{Tr}(\rho A)
	\end{equation}
	which indeed satisfies the KMS condition for the correlation function,
	\begin{equation}
	    F(A,B,t+i\beta) = \omega((\alpha_t B)A) = \omega(A \alpha_t B) = F(A,B,t)
	\end{equation}
	via cyclicity of the trace operation, meaning that this state is analytic on the boundary of the strip $0<Im(t)< \beta$. This constitutes the cyclic vector $\ket{\Omega}$ as a thermal equilibrium state. Jacobson has emphasized \cite{Jacobson2016} that the vacuum in any quantum field theory can be expressed as a thermal density matrix when restricted to a causal diamond. This may suggest that our cyclic vector may be related to a vacuum of some type, a point we will return to in a later section. 
	
	Due to the fact that the time flow emerges from the modular group of a specific KMS equilibrium state, $\omega$, Connes and Rovelli \cite{Connes1994} have hypothesized that physical time flow has thermodynamic origin in the sense that the physical time flow, $\alpha_t$, depends upon the thermal state of the system, $\omega$. This is known as the thermal time hypothesis. They were unable to connect such thermal modular time flow to a suitable metric coordinate for time as in relativity theory. It is within the current model that this connection can be made, as the (modular) thermal time evolution in conformal quantum mechanics is deeply connected to the conformal radial Killing symmetries of a causal diamond in Minkowski space as known from the relation $R=i\chi$. The fact that a conformal radial Killing field is related to quantum $operators$ as opposed to quantum states is a reflection of the fact that it is really the operators, and the algebraic relations they obey, that define a theory as is the philosophy of algebraic quantum field theory \cite{Hollands2015}. This is a step toward bridging the gap between thermal modular flow of time and dynamic geometric time flow of a causal diamond. 
	
Entropy of entanglement can be used as a measure of entanglement for a bipartite quantum system such as the vector $\ket{\Omega}$ by computing the von Neumann entropy for the reduce density matrix of the subsystems. Observing that the entanglement entropy is non zero results in a finite von Neumann entropy, meaning that the reduced density is a mixed state thus indicating entanglement for the bipartite state. We can compute the entanglement entropy of the cyclic vector $\ket{\Omega}$ as,
    \begin{equation}
        \begin{array}{c}
            S_{\Omega} = -Tr_a(\rho_{\Omega}^a \; ln \rho_{\Omega}^a) = -Tr_b(\rho_{\Omega}^b \; ln \rho_{\Omega}^b) \\\\
           = -N^2(\beta)\sum_{n=0}^{\infty} e^{-\beta n}(-\beta n + ln (N^2(\beta)) = \dfrac{N^2(\beta) \beta e^{-\beta}}{(1-e^{-\beta})^2} - 2\dfrac{N^2(\beta)}{1 -e^{-\beta}}ln(N(\beta))
        \end{array}
    \end{equation}
 	indicating entanglement for the vector $\ket{\Omega}$. Plugging in the normalization $N(\beta) = (1-e^{-\beta})^{1/2}$ and using the relation $\beta = \dfrac{1}{T} = \pi \alpha$, we obtain 
 	 \begin{equation}
        S_{\Omega} =  \dfrac{\pi \alpha }{(e^{\pi \alpha}-1)} - ln(1-e^{-\pi \alpha})
    \end{equation}
 	 which is an expression for the entanglement entropy of the cyclic vector representing two causal diamonds in terms of the length, $\alpha$ of one diamond, minus a logarithmic corrective term. This is similar to a well known logarithmic correction in holography theory \cite{Carlip2000}. This is expected since the quantum theory represents only $(1+1)$ components of the full $(3+1)$ causal diamond.
 	 
	In summary, to generate the conformal radial symmetries of two causally disconnected diamonds, one need define two sets of commuting conformal quantum operators. The non-locality of the entanglement in algebraic quantum field theory is natural in a conformal model where the causal diamonds themselves are seen to emerge from an algebraic structure of conformal quantum operators that compose von Neumann algebras. We will see in the next section how the entanglement of the model is quantified using the modular conjugation operator $J$.
	
\section{A Concurrence Measurement for Entanglement using Modular Operators}

	A well known way to measure entanglement is Wooter's Concurrence \cite{Wooters1997}. 
The entanglement of formation of a generic bi-partite state $\rho$ (quantum density operator) is related to the concurrence of that state $C(\rho)$ as follows:
\begin{equation*}
E_{Formation}(\rho)=H_{bin}\left( \dfrac{1+\sqrt{1+C(\rho)}}{2} \right)
\end{equation*}
where $H_{bin}$ is the Shannon binary entropy.
 One can define the concurrence for a bi-partite state using the anti-unitary modular conjugation operator as \cite{Chatterjee2021}
	
	\begin{equation} 
		C(\ket{\psi}) = | \bra{\psi} J \ket{\psi } | 
		\label{Concurrence}
	\end{equation} 
	The concurrence relation (\ref{Concurrence}) has given a physical meaning to the modular conjugation operator as a quantitative measure of entanglement for a bi-partite state. It is known that one needs an anti-linear, anti-unitary operator as the key driving force behind the entanglement of formation of bipartite systems and it is now understood in this context as the modular conjugation operator $J$.

	The most convenient basis to use for an entanglement calculation here is that of quantized rotational states, for these are discrete, normalized states that complete the Hilbert space. For an entangled rotational state $\ket{\Psi} = \alpha \ket{n} \otimes  \ket{m} + \beta \ket{m}\otimes\ket{n}$, and for $n \neq m$, we have a concurrence given by the modular conjugation operator $J$ as
	\begin{equation}
    	C(\ket{\Psi})=| \bra{\Psi} J \ket{\Psi } |= 2|\alpha \beta|. 
    \end{equation}
Clearly when $\alpha = \beta = \dfrac{1}{\sqrt{2}}$ we have a maximally entangled state of $C = 1$. The nonzero entanglement again is interpreted to arise from the active separation of causal compliments, $\mathcal{A}$ and $\mathcal{A}^{'}$. This gives a measure of the entanglement between two spacelike separated causal diamonds in Minkowski space as was achieved by representing the radial and temporal dimensions of each diamond as a set of conformal quantum mechanical operators that generate rotational states in a (0+1) conformal field theory. 
	
	\section{Thermal Structure of Conformal Quantum Mechanics}

	It was mentioned above that the thermal time flow can be derived from the modular automorphism group for the current theory, $\alpha_t A = \Delta^{it}A\Delta^{-it} = e^{it\beta(N_a \otimes I - I\otimes  N_b)}Ae^{-it\beta(N_a \otimes I - I\otimes  N_b)}$. The modular flow induced by the thermality of the system can be clearly viewed in conformal quantum mechanics. This is due to the fact that the quantization of time was hinted at in the original observation of conformal symmetry in a $(0+1)$ field theory \cite{Alfaro1976}. This view leads to the quantization of states $\ket{t}$ labeled by a time variable $t$.  These states are in fact the Fourier transform of the energy eigenkets of the Hamiltonian,
    \begin{equation}
        \ket{E} = 2^{r_0}E^{\frac{1}{2}-r_0} \int_{-\infty}^{\infty} dt e^{-iEt}\ket{t}
    \end{equation}
    that form a continuous, orthonormal basis. Indeed, one can construct quantized time states from the rotational vacuum through a complex time translation \cite{Alfaro1976},
	
	\begin{equation} 
		\ket{t} = e^{iHt}\ket{t=0} = \frac{\Gamma^{\frac{1}{2}}(2r_0)}{2^{2r_0}}e^{(\alpha+it)H} \ket{n=0}
	\end{equation}
    where $\alpha$ is half the length of the causal diamond. One may interpret this to mean that quantized time emerges from the thermal evolution of the vacuum. The quantization of time was thoroughly discussed in \cite{Caldirola1956, Farias2010, Caldirola1980, Albanese2004, Olkhovsky2007, Yang1947} among many other works, though it is currently unclear whether the time states $\ket{t}$ corresponds to states of the hypothetical "chronon", which is a discrete unit of time of $6.266*10^{-24}s$. Furthermore, it has been shown that these are thermal states by observing the two point function for "$\tau$" time states of the quantum operator $S$ \cite{Arzano2021}, 
	
	\begin{equation} 
		G(\tau_1, \tau_2) = \braket{\tau_1| \tau_2}  = \frac{\Gamma (2r_0) \alpha^{2r_0}}{(4isinh(\frac{\tau_1 - \tau_2}{2}))^{2r_0} }
	\end{equation}
at a temperature of $\dfrac{1}{2\pi}$ and conformal weight $\Delta = r_0$. It was mentioned in \cite{Arzano2020,Gupta2017} that there is a direct connection between this thermal two-point function of conformal quantum mechanics and the one associated to scalar field along the world line of a static diamond observer in spacetime. Note that our thermal cyclic vector (\ref{cyclic}) is related to the temporal vacuum of \cite{Arzano2021}, $\ket{\tau = 0} = \sum_n (-1)^n \ket{n} \otimes \ket{n}$, coinciding with arguments that the vacuum is thermal when restricted to a causal diamond \cite{Jacobson2016}. Other thermal effects of conformal quantum mechanics and thermal time flow were investigated in \cite{Martinetti2003, Martinetti2009}. This link also provides a possible path to the Reeh-Schlieder theorem for conformal quantum mechanics, which will be made more precise in future work. Further, this thermal structure supports the interpretation that spacetime and causality may emerge from thermality. The physical interpretation can be made more clear by noticing that the creation operator for time states, $\mathcal{O}(t)$, is related to the creation operator for rotational states, $L_+$, as in \cite{Jackiw2012} 
	
	\begin{equation}
		\begin{array}{cc}
			\mathcal{O}(t) = N(t)e^{\omega(t)L_+} \\\\
			 N(t) = \left(\Gamma(2r_0)\right)^{1/2} \left(\dfrac{\omega(t)+1}{2}\right)^{2r_0}, \\\\
			\omega(t)=\dfrac{\alpha+it}{\alpha-it} = e^{i\theta}, \; t=\alpha \; tan\theta/2
		\label{O}\end{array}
	\end{equation}
 suggesting that the construction of a quantum time variable labeled by the ket $\ket{t}$ may be interpreted as a form of abstract angular momentum scaled by some temperature. 
 
 Using the current construction of von Neumann algebras and modular operators of conformal quantum mechanics from section III, we can determine the modular structure of time states created from the commutant algebra for a second causal diamond. Let $O_a(t) = e^{(\alpha+it)H_a}$ such that 
   
\begin{equation}
    \ket{t} = \ket{t_a} \otimes \ket{t_b} = \biggl(O_a(t) \otimes O_b(t)\biggr) \ket{0} \otimes \ket{0}
\end{equation}
Our task is to determine the form of the operator $O_b(t)$ such that it composes the commutant of $\mathcal{A}_a$. Using the above relations, we find that, 

\begin{equation}
    \begin{array}{cc}
        J (O_a \otimes I) J \ket{t} = J (e^{(\alpha +it)H_a}\otimes I) J (\ket{t_a} \otimes \ket{t_b})\\\\
        =J (e^{(\alpha +it)H_a} \otimes I) (\ket{t_b} \otimes \ket{t_a}) = J (e^{(\alpha+it)H_a}\otimes I)(O_b(t) \otimes O_a(t) \ket{0} \otimes \ket{0}) \\\\
        =(I\otimes e^{(\alpha -it)H_b}) (O_a(t) \otimes O_b(t) \ket{0} \otimes \ket{0}) = (I\otimes O_b(t)) \ket{t}. 
    \end{array}
\end{equation}
with $O_b(t) = e^{(\alpha-it)H_b}$. Indeed, in a similar fashion to equations (31, 32) we have the commutant structure $J(O_a(t)\otimes I) J=(I\otimes O_b(t))$. Therefore we find that the state $\ket{t_b}$ is created from the vacuum at an inverse temperature $\dfrac{1}{\alpha-it}$ with opposite complex time $-it$ as opposed to $\ket{t_a}$ which is created from the vacuum at the inverse temperature $\dfrac{1}{\alpha+it}$ and complex time $it$. This therefore suggests that such a quantum time generated from those states is flowing in opposite complex directions. If one associates the quantum time states as representing the time in a causal diamond, then this would suggest opposite temperatures and thus complex flows of time for causally disconnected regions. The concept of negative temperature in the context of causal diamonds was discussed thoroughly in \cite{Jacobson2019, Visser2019}. However, it is currently unclear how to relate the time kets, $\ket{t}$, and their eigenvalues \cite{Chamon2011}, $-r_0 \omega$, to physical time, $t$.

The modular group flow, $\alpha_t A = \Delta^{it} A \Delta^{-it}$, allows us to define a notion of thermal time flow in the current model. We therefore find that the ordinary time evolution of a quantum observable, $\gamma_t A = e^{itH/\hbar}Ae^{-itH/\hbar}$, can be equated with the modular time flow, $\alpha_t A = \Delta^{it} A \Delta^{-it}$, when $H = \hbar \beta (N_a \otimes I - I\otimes  N_b) $ thus giving $\gamma_t A = \alpha_t A$ \cite{Connes1994}. The thermal time hypothesis, namely that the time flow is determined by the thermal state, is then supported for a theory including thermal states of quantized time. It is a topic of future work to further investigate the application of modular operators to time states $\ket{t}$.

Indeed, one can approximately describe the dynamics of an eternal black hole in a maximally extended AdS by considering two entangled conformal field theories on each boundary, which is the theory of thermo-field dynamics \cite{Maldacena2003, Israel1976}. In that framework, a thermal density operator in the conformal field theory corresponds to a black hole geometry. The formalism developed in this work can be understood to be associated with the formalism of thermo-field dynamics, as our cyclic vector is the thermo-field double for two conformal field theories in $(0+1)$ dimensions.

	\section{Discussion}
	We have elucidated how the entanglement of two causally closed regions emerges from modular operator structure developed for conformal quantum field theory in (0+1) dimensions. This was achieved by constructing local von Neumann algebras for conformal quantum mechanics, which is known to be holographically dual to a causal diamond spacetime structure. The representation of two von Neumann algebras is then studied using Tomita-Takesaki theory, and connections between two causal diamonds are made by representing each of them as local conformal quantum mechanical algebras. Entanglement between the quantum states representing each diamond was quantified using Tomita-Takesaki modular operators through entropy and Wooter's Concurrence as defined in \cite{Chatterjee2021}. The entanglement entropy of the cyclic vector of conformal quantum mechanics was found to be proportional to the length of the causal diamond with a logarithmic correction, which is expected for a holographic conformal theory. A physical interpretation is provided by seeing the radial conformal Killing field in terms of the quantum field as in equation (\ref{killingfieldquantumfield}). This provides us with a simplified model for understanding how spacetime and causality may emerge from the entanglement entropy of causal diamonds, temperature, and quantum states. 
	
	It was also mentioned how to understand the thermal time hypothesis in terms of the quantized states found in conformal quantum mechanics. By examining the algebraic structure of the quantized time states in conformal quantum mechanics, two causally disconnected regions were found to be at opposite temperatures and thus opposite flows of complex time. The states were shown to be KMS states and thus generated the modular automorphisms from which local time flow emerges. It is of interest to us to understand better the relation between quantum time in conformal quantum mechanics and physical time flow in a causal diamond. Perhaps at the scale when quantum gravitational effects become relevant, the current model suggests at such a scale that space is not a fundamental construct, but rather temperature is the fundamental concept from which classical spacetime, causality and gravitation emerge. We plan to investigate further the bulk geometric spacetime structure that emerges from the correspondence in a future work. 
	
	\section{Appendix: Von Neumann Algebras and Tomita-Takesaki Modular Operators}
	
	The general references for this section are \cite{KR1983,Tak1979,Bratteli1987}.
	
	\noindent\textit{\textbf{Normed Algebra:}} Consider an algebra  $A \in \mathcal{A}$ over $\mathbb{C}$. A normed algebra has a Norm Map: $A \rightarrow ||A|| \in \mathbb{R}^+$ such that
	
	$||A|| > 0$
	
	$||A|| = 0 \; \text{iff} \; A=0$
	
	$\alpha \in \mathbb{C},  ||\alpha A||= |\alpha| ||A||$
	
	$||A+B|| \leq ||A|| + ||B||$
	
	$||AB|| \leq ||A|| \; ||B||$
	
	\bigskip
	
	\label{BanachAlg}
	\noindent\textit{\textbf{Banach Algebra:}} Let $\mathcal{A}$ be an algebra. A normed algebra has a norm map: $\mathcal{A} \rightarrow \mathbb{R}^+ , \,\,\, A \rightarrow ||A|| \in  \mathbb{R}^+ , \forall A \in \mathcal{A} $ such that
	
	\medskip
	
	$||A|| \geq 0,$
	
	$||A|| = 0 \iff A=0, $
	
	$\alpha \in \mathbb{C} , ||\alpha A || = |\alpha| \, ||A||, $
	
	$|| A+B|| \leq ||A|| + ||B||, $
	
	$||AB|| \leq ||A|| \, ||B|| .$
	
	\bigskip
	
	\noindent A Banach algebra is a complete normed algebra (complete in the norm map): Consider a Cauchy sequence $A_1, A_2, ... \in \mathcal{A}$. For any $\epsilon >0, \exists$ an integer $N$ such that $\forall$ natural numbers $m,n > N$, $||A_m -A_n|| < \epsilon$ (a sequence whose elements become arbitrarily close to each other as the sequence progresses). 
	By introducing a "distance" metric $d(A,B)=||A-B||$, once induces a topology on $\mathcal{A}$ where a neighborhood $U$ in  $\mathcal{A}$ is given by $U(A,\epsilon)=\{B; \; B \in \mathcal{A}, \; d(A,B) < \epsilon, \; \epsilon >0\}$. A \textit{metric space} $(\mathcal{A}, d)$ in which every Cauchy sequence converges to an element in $\mathcal{A}$ is called \textit{complete} in the standard norm.
	\textit{Counterexample}: Rational numbers $(p/q) \in \mathbb{Q}$ are not complete. The Cauchy sequence $x+0=1, x_{n+1}=\dfrac{x_n +2/x_n}{2}$ converges to the irrational number $\sqrt{2}$. 
	
  Furthermore, for example, let $\mathcal{A} = C^0(X, \mathbb{C}) \equiv$ complex valued continuous functions $f$ on a compact space $X$ (a bounded region): $f: X \rightarrow \mathbb{C}$.
	\medskip
	
	Let $x \in X, \;\;\ f,g \in \mathcal{A}$
	
	$(f+g)(x) =f(x) +g(x)$ (Addition)
	
	$(\alpha f)(x) = \alpha f(x)$ (scalar multiplication)
	
	$\Big( fg \Big) (x) =f(x)g(x)$ (Algebra product)
	
	$||f|| = \displaystyle\sup_{x\in X} |f(x)|$ (norm)
	
	\bigskip
	
	\bigskip \label{CStarAlg}
	\noindent\textit{\textbf{$\bm{C^*}$-Algebra:}} A $\bm{C^*}$-algebra $\mathcal{A}$ is a Banach algebra with an involutive map $ ^*: \mathcal{A} \rightarrow \mathcal{A} , \,\,\, A \rightarrow A^* \,\,\, \forall A \in \mathcal{A} , \lambda \in \mathbb{C}$ such that
	
	\medskip
	
	$(A^*)^*=A$
	\medskip
	
	$(AB)^*=B^* A^*$
	\medskip
	
	$(\lambda A)^*= \bar{\lambda} A^*$
	\medskip
	
	$ (\lambda  A + \alpha B )^* = \bar{\lambda } A^* +\bar{\alpha}  B^*  $ (anti-linear)
	(so far, we have a $*$-algebra)
	\medskip
	
	$||A^*|| = ||A|| $ (norm condition for involutive Banach algebra),
	\medskip
	
	$||A A^*|| = ||A|| \, ||A^*|| = ||A^* A|| = ||A^*|| \, ||A|| =||A||^2$ ($C^*$ condition for $\bm{C^*}$-Algebra).
	
	\bigskip
	
	\noindent If $A^* = A $, this element is called \textit{self-adjoint}. If $A^*A = AA^* =I$ this element is called \textit{unitary}.
	
	\bigskip
	
For example, let $\mathcal{A} = C^0(X, \mathbb{C}) \equiv$ complex valued continuous functions $f$ on a compact space $X$ (a bounded region). From above, we know that this is a Banach algebra. We  know introduce an involution map through complex conjugation:
	\medskip
	
	$f^*: X \rightarrow \mathbb{C}, \;\;\; f^* \in \mathcal{A}$
	
	\medskip
	
	$f^*(x) := \overline{f(x)}, \; \forall x \in X$ (Involution)
	\medskip
	
	$||f^*|| = \displaystyle\sup_{x\in X} |f^*(x)|=\sup_{x\in X} |\overline{f(x)}| =\sup_{x\in X} |f(x)|= ||f||$ (norm condition)
	
	\bigskip
Furthermore, consider complex $n \times n$ matrices $A$. Here, we have
	
	\medskip
	$A^* := A^{\dagger}$ (involution is the adjoint operation)
	
	\medskip
	$||A|| := \sqrt{Tr(AA^{\dagger})}$ (norm is the square root of trace)
	
	\medskip
	\noindent The algebra is given by commutative matrix addition $A+B$ and non-commutative matrix multiplication $AB \neq BA$.
	
	\bigskip
	
	\noindent\textit{\textbf{Bounded Linear Operators $\mathcal{B}(\mathcal{H})$ on a Hilbert Space}:} Let $\mathcal{H}$ be a Hilbert space. A bounded linear operator $A \in \mathcal{B}(\mathcal{H})$ acts on the Hilbert space $A: \mathcal{H} \rightarrow  \mathcal{H}$ such that
	
	\medskip
	
	$A(\psi + \phi)=A(\psi)+A(\phi), \;\; \forall \psi,\phi \in \mathcal{H}$ (linearity)
	
	\medskip
	
	$\exists$ a positive real number $c < \infty$ such that $||A(\psi)|| \leq c ||\psi||, \;\; \forall \psi \in \mathcal{H}$ (bounded)
	
	\medskip
	
	$||A|| =  \displaystyle\sup_{\psi \in \mathcal{H}, ||\psi|| =1} \{||A(\psi)|| \} $ (norm of an operator)
	
	\medskip
	
	\noindent For all $A \in \mathcal{B}(\mathcal{H}), \;\exists$ a unique element called the \textit{adjoint operator} $A^{\dagger} \in \mathcal{B}(\mathcal{H})$ such that (using the Hilbert space inner product $\langle \cdot , \cdot \rangle$
	
	\medskip
	
	$\langle A^{\dagger}(\psi), \phi \rangle = \langle \psi, A(\phi) \rangle $
	
	\medskip
	
	$(\alpha A + \lambda B)^{\dagger} = \overline{\alpha} A^{\dagger} + \overline{\lambda} B^{\dagger}$
	
	\medskip
	
	$(A B)^{\dagger} = B^{\dagger} A^{\dagger} $
	
	\medskip
	
	$(A^{\dagger})^{\dagger} = A $
	
	\medskip
	
	$||A^{\dagger}A|| = ||A A^{\dagger}|| =||A||^2 $
	
	\medskip
	
	$||A^{\dagger}|| = ||A||$
	
	\bigskip

	\bigskip
	
	The \textit{$\bm{C^*}$-Algebra} features can be given to $\bm{\mathcal{B}(\mathcal{H})}$ via the following
	
	\medskip
	Banach Norm $\longrightarrow$ Operator Norm
	
	\medskip
	Involution $*$-operator $\longrightarrow$ Adjoint operation $^\dagger$
	
	\medskip
	$||A^{*}||=||A^{\dagger}|| = ||A||$ (Norm condition)
	
	\medskip
	$||A A^*|| = ||A A^{\dagger}|| =||A|| \; ||A^{\dagger}|| = ||A||^2$ ($C^*$ condition for $\bm{C^*}$-Algebra).
	
	\bigskip
	
	\noindent\textit{\textbf{$\bm{^*}$-Homomorphism $\bm\xi$:}} Let $\mathcal{A}$  and $\mathcal{B}$ be $\bm{C^*}$-algebras. A  $\bm{^*}$-homomorphism $\xi$ is a mapping $\xi:\mathcal{A} \rightarrow \mathcal{B}$  that preserves the algebraic and $\bm{^*}$ structures of $\mathcal{A}$. That is, $\forall A,A_1,A_2 \in \mathcal{A}$
	
	\medskip
	
	$\xi( A_1 + \ A_2)=  \xi(A_1) +  \xi(A_2)$ (linearity)
	
	\medskip
	
	$\xi(A_1 A_2)= \xi(A_1)\xi(A_2)$ (homomorphism)
	
	\medskip
	
	$\xi(A^*)=\xi(A)^*$ ($^*$-preserving)
	
	\medskip
	
	\noindent In general, $\bm\xi$ is norm decreasing, i.e. $||\xi(A)|| \leq ||A||$. If  $\bm\xi$ is an $\bm{^*}$-\textit{isomorphism} (one-to-one, onto), then it is norm preserving, $||\xi(A)|| = ||A||$. A $\bm{^*}$-\textit{automorphism} is a $\bm{^*}$-isomorphism from a $\bm{C^*}$-algebra to itself, i.e. $\xi:\mathcal{A} \rightarrow \mathcal{A}$.
	
	\bigskip
	
	\noindent\textit{\textbf{Representation $\bm\pi$ of a $\bm{C^*}$-Algebra:}} A representation of a $\bm{C^*}$-algebra $\mathcal{A}$ is a pair $(\mathcal{H}, \pi)$ where $\mathcal{H}$ is a Hilbert space and $\pi$ is a $\bm{^*}$-Homomorphism from $\mathcal{A}$ to $\mathcal{B}(\mathcal{H})$, $\pi:\mathcal{A} \rightarrow  \mathcal{B}(\mathcal{H})$. If $\pi$ is an $\bm{^*}$-isomorphism, it is called \textit{faithful} such that
	\medskip
	
	$\text{ker} \; \pi = \{0\}$ (faithful)
	
	\medskip
	
	$||\pi(A)|| = ||A||, \forall A \in \mathcal{B}(\mathcal{H})$ (norm preserving)
	
	\medskip
	
	$||\pi(A)|| > 0, \forall A > 0$ (positive)
	
	\medskip
	\noindent A subset $S \subset \mathcal{H}$ is called \textit{\textbf{invariant}} under $\mathcal{A}$ if $\pi(\mathcal{A})S := \{\mathcal{A}\psi | A\in \mathcal{A}, \psi \in \mathcal{H}\} \subset S$
	
	\medskip
	\noindent A representation $\pi$ is called \textit{\textbf{irreducible}} if the only subspace of $\mathcal{H}$ invariant under $\pi(\mathcal{A})$ are $\{0\}$ and $\mathcal{H}$.
	
	\medskip
	\noindent An operator ${A \in \mathcal{B}(\mathcal{H})}$ is called \textit{\textbf{dense}} if its range exists only in the dense subspace of the Hilbert space $\mathcal{D} \in \mathcal{H}$.
	
	\bigskip 
	
	\noindent\textit{\textbf{Cyclic Vector  $\; \Omega \in \mathcal{H}$:}} A vector $\; \Omega \in \mathcal{H}$ is called a cyclic vector for a set of bounded operators $\mathcal{B}(\mathcal{H})$ if $\{A \; \Omega | A \in \mathcal{B}(\mathcal{H})\}$ is \textbf{dense} in the whole $ \mathcal{H}$.
	
	\bigskip
	
	\noindent\textit{\textbf{Cyclic Representation $\bm\pi$ of a $\bm{C^*}$-Algebra:}} A cyclic representation of a $\bm{C^*}$-algebra $\mathcal{A}$ is a triple $(\mathcal{H}, \pi, \Omega)$ where $(\mathcal{H}, \pi)$ is a representation of $\mathcal{A}$ and $\Omega \in \mathcal{H}$ is cyclic in the representation $\pi$.

	\bigskip
	
	\noindent \textit{\textbf{Linear Functionals (and States):}} Given an algebra $\mathcal{A}$, a linear functional (or state) on $\mathcal{A}$ (a scalar valued 'function' on $\mathcal{A}$)  is a map 
	$\omega:\mathcal{A} \rightarrow \mathbb{C}$ such that $\omega(\lambda A +\beta B)=\lambda \omega(A) +\beta \omega(B)$. Furthermore,
	
	\medskip
	
	If $\mathcal{A}$ is a $^*$-algebra, $\omega$ is called a \textbf{positive functional} if $\omega(A^* A) \geq 0, \forall A \in \mathcal{A}$
	
	\medskip
	
	A \textit{\textbf{state}} $\omega$ is  a \textbf{positive functional} with $\omega(A^*)=\overline{\omega(A)}$.
	
	\medskip
	
	If $\mathcal{A}$ is a $^*$-algebra, and $\omega_i$ are \textbf{positive functionals} then for 
	$\lambda_i \in \mathbb{R}^+, \sum_i \lambda_i = 1$, one may construct a \textit{positive convex linear functional as} $\omega = \sum_i \lambda_i \omega_i$.
	
	\medskip
	
	If $\mathcal{A}^{sa}$ is self-adjoint subspace, i.e. $A^* = A$ and $\omega$ is a \textbf{positive functional}, then  $\omega:\mathcal{A}^{sa} \rightarrow \mathbb{R}$ (real scalars)
	
	\medskip
	
	If $\mathcal{A}$ is a $\bm{C^*}$-algebra, every positive functional $\omega$ is continuous.
	
	\medskip
	
	Let $\mathcal{A}$ be a $\bm{C^*}$-algebra. $\forall A \in \mathcal{A}, \exists$ a positive functional $\omega_A$ such that, $||\omega_A||=1$ and $\omega_A(A^* A) = ||A||^2$.

	\bigskip
	
	\noindent \textit{\textbf{GNS Construction}-Representation from a State:}
	Let  $\mathcal{A}$ be a $\bm{C^*}$-algebra (of observables). Let $\omega$ be a functional (state) on  
	$\mathcal{A}$. A representation of $\mathcal{A}$ constructed from $\omega$  is a pair $(\mathcal{H}, \pi_{\omega})$ where given the inner product on $\mathcal{H}$, $\left\langle \;\; | \;\; \right\rangle: \mathcal{H} \times \mathcal{H} \rightarrow \mathbb{C}$ and a cyclic vector $\Omega_\omega$ with respect to $\pi_{\omega}$ $\bigl( \pi_{\omega}(\mathcal{A}) |\left. \Omega_\omega \right\rangle$ is dense in $\mathcal{H}\bigr)$, one has the action of operators on $\mathcal{H}$ such that
	\begin{equation*}
		\omega(A) := \dfrac{\left\langle \Omega_\omega |\pi_{\omega}(A) \Omega_\omega \right\rangle}{\left\langle \Omega_\omega | \Omega_\omega \right\rangle}
	\end{equation*}
	\noindent Remark: This method requires one to choose a preferred representation $\pi_{\omega}$. In quantum mechanics, the Stone von Neumann theorem states that all representations of the Weyl algebra are unitarily equivalent to the Schr\"{o}dinger representation so this is not a problem. In Minkowski QFT, Poincar\'{e} invariance chooses an appropriate cyclic vector (vaccuum state). The choice of a representation becomes an issue in QFT on curved spacetime manifolds.
	
	\bigskip
	
	\noindent \textit{\textbf{Von Neumann Algebra:}} Consider a  $C^*$-algebra $\mathcal{B(H)} =\{A\}$ of bounded linear operators on a Hilbert space, $ A: \mathcal{H} \rightarrow \mathcal{H}$. Let $\mathcal{C}$ be a subset of $\mathcal{B(H)}$. An operator $A \in \mathcal{B(H)} $ belongs to the commutant $\mathcal{C}'$ of the set $\mathcal{C}$ $\iff$   $AC =CA, \,\,\, \forall C \in \mathcal{C}$. A von Neumann algebra $\mathcal{A}$ is a unital $C^*$-subalgebra of $\mathcal{B(H)}$ such that $\mathcal{A}'' = \mathcal{A}$. Consider a von Neumann algebra $\mathcal{A} \subset \mathcal{B(H)}$. A von Neumann algebra in standard form is one where there exists an element $| \Omega \rangle \in \mathcal{H}$ which is both cyclic (operating on $| \Omega \rangle $ with elements in  $\mathcal{A}$ can generate a space dense in $\mathcal{H}$) and separating (if $A | \Omega \rangle = 0$, then $A=0$).

	\bigskip

	\noindent \textit{\textbf{Tomita Takesaki Modular Operators:}} Consider a  von Neumann algebra $\mathcal{A} \subset \mathcal{B(H)}$ in standard form with a cyclic and separating vector $| \Omega \rangle \in \mathcal{H}$. Let $S:\mathcal{H} \rightarrow \mathcal{H}$ be a anti-unitary operator defined by $S A  | \Omega \rangle = A^* | \Omega \rangle $. Let the closure of $S$ have a polar decomposition given by $S=J \Delta ^{\frac{1}{2}} =\Delta^{-\frac{1}{2}} J$, where $J$ is called the modular conjugation operator and $\Delta$ is called the modular operator. $J$ is anti-linear and anti-unitary whereas $\Delta$ is self-adjoint and positive. Furthermore, the following relations hold:
	
	1. $J \Delta ^{\frac{1}{2}} J = \Delta ^{-\frac{1}{2}}$
	
	2. $ J^2 =I, \,\,\, J^{*} = J$
	
	3. $ J |\Omega \rangle  = |\Omega \rangle  $
	
	4. $ J \mathcal{A} J = \mathcal{A}' $
	
	5. $ \Delta = S^{*} S$
	
	6. $ \Delta |\Omega \rangle  = |\Omega \rangle $
	
	7. $ \Delta^{it} \mathcal{A} \Delta^{-it} = \mathcal{A} $ (one parameter-$t$ group of automorphisms of $\mathcal{A})$
	
	8. If $ \omega (A) = \langle \Omega | A \Omega \rangle $, $ \forall A \in \mathcal{A}$,  then $\omega$ is a KMS (Kubo-Martin-Schwinger) functional (state) on $\mathcal{A}$ with respect to the automorphism of 7.
	
	9.$|\Omega \rangle $  is cyclic for $\mathcal{A}$ if and only if $|\Omega \rangle $  is separating for $\mathcal{A}'$
	
	\section{Acknowledgements}
	We would like to thank Michele Arzano for helpful discussions.

\end{document}